\documentclass{article}

\usepackage[utf8]{inputenc} 
\usepackage[T1]{fontenc} 
\usepackage{hyperref} 
\usepackage{url} 
\usepackage{booktabs} 
\usepackage{amsfonts} 
\usepackage{amsmath} 
\usepackage{nicefrac} 
\usepackage{microtype} 
\usepackage{fancyhdr} 
\usepackage{graphicx}
\usepackage{cite}
\usepackage{doi}
\usepackage{array}
\usepackage{arxiv} 
\usepackage{authblk}
\usepackage{tabu}
\usepackage{pdfpages}

\graphicspath{{fig/}}

\title{Interpretable models for extrapolation in scientific machine learning}

\author[1]{Eric S. Muckley}
\author[1]{James E. Saal}
\author[1]{Bryce Meredig}
\author[2]{Christopher S. Roper}
\author[2]{John H. Martin}

\affil[1]{Citrine Informatics, Redwood City, 94063, CA, USA}
\affil[2]{HRL Laboratories, Malibu, 90265, CA, USA}

\begin{document}
\maketitle

\fancypagestyle{firststyle}
{
  \fancyhf{}
  \fancyhead[C]{\footnotesize DISTRIBUTION STATEMENT A (Approved for Public Release, Distribution Unlimited)}
   \fancyfoot[C]{\footnotesize DISTRIBUTION STATEMENT A (Approved for Public Release, Distribution Unlimited)}
   \renewcommand{\headrulewidth}{0pt} 
}

\thispagestyle{firststyle}

\fancyhead[EC, OC]{\footnotesize DISTRIBUTION STATEMENT A (Approved for Public Release, Distribution Unlimited)}
\fancyfoot[EC, OC]{\footnotesize DISTRIBUTION STATEMENT A (Approved for Public Release, Distribution Unlimited)}

\begin{abstract}

Data-driven models are central to scientific discovery. In efforts to achieve state-of-the-art model accuracy, researchers are employing increasingly complex machine learning algorithms that often outperform simple regressions in interpolative settings (e.g. random k-fold cross-validation) but suffer from poor extrapolation performance, portability, and human interpretability, which limits their potential for facilitating novel scientific insight. Here we examine the trade-off between model performance and interpretability across a broad range of science and engineering problems with an emphasis on materials science datasets. We compare the performance of black box random forest and neural network machine learning algorithms to that of single-feature linear regressions which are fitted using interpretable input features discovered by a simple random search algorithm. For interpolation problems, the average prediction errors of linear regressions were twice as high as those of black box models. Remarkably, when prediction tasks required extrapolation, linear models yielded average error only 5\% higher than that of black box models, and outperformed black box models in roughly 40\% of the tested prediction tasks, which suggests that they may be desirable over complex algorithms in many extrapolation problems because of their superior interpretability, computational overhead, and ease of use. The results challenge the common assumption that extrapolative models for scientific machine learning are constrained by an inherent trade-off between performance and interpretability.

\end{abstract}


\section*{Introduction}

Machine learning has become a principal catalyst for scientific discovery, particularly in the design of novel functional materials~\cite{iwasaki2019identification, wei2019machine}. In efforts to build predictive models with state-of-the-art performance, researchers are employing increasingly complex black box algorithms, including large-scale ensembles and deep neural networks, because of their ability to approximate high-dimensional response surfaces with arbitrary precision~\cite{agrawal2019deep}. Recent availability of low-cost data storage and computing resources has unlocked model architectures capable of handling large numbers (10$^{5}$ – 10$^{7}$) of input features~\cite{maniruzzaman2018accurate, tetko2016development}, enabling the development of deep learning models such as Elemnet~\cite{jha2018elemnet} and SchNet~\cite{schutt2018schnetpack} which can learn feature encodings directly from the elemental compositions of input materials. While the complexity of these models often leads them to outperform traditional regression techniques by standard cross-validation scoring metrics, they also suffer from notable disadvantages~\cite{guidotti2018survey, yang2018deep}.


Increases in model complexity are generally accompanied by corresponding decreases in model portability and usability by non-experts. Many state-of-the-art black box algorithms require significant computational resources and hyperparameter optimization during training, which limits their usefulness in edge computing environments and necessitates experienced practitioners for managing model serialization, programming environments and version control systems, and input data compatibility~\cite{baier2019challenges, paleyes2020challenges}. Growing complexity also inherently limits model interpretability by humans~\cite{murdock2020domain, butler2018machine}, which increases the likelihood of model overfitting, reduces researcher trust in predictions, and creates difficulty in troubleshooting~\cite{rudin2019stop}. Perhaps most importantly, poor model interpretability impedes domain expert intuition by obscuring natural underlying patterns that often guide researchers toward novel insights and new physics~\cite{wagner2016theory, iwasaki2019identification,lei2021aggressively}.

The use of simple interpretable models has garnered renewed attention amid mounting evidence that the trade-off between model performance and interpretability is often overstated~\cite{iwasaki2019identification, rudin2019stop}. Interpretability aids in diagnosis of model biases, management of multi-objective trade-offs, and mitigation of unexpected results~\cite{azodi2020opening, mikulskis2019toward, doshi2018considerations, guidotti2018survey,xiang2021physics}, which are essential considerations in the design of materials and chemicals for novel drugs, electronics, catalysts, and alloys~\cite{wei2019machine, mueller2016machine}. The primary challenge of materials informatics, accurate prediction of the physical properties of a material from its chemistry or other known characteristics, relies on the discovery of interpretable physics-informed input features: empirically or theoretically derived vectors of known quantities which can be used to predict the value of a target property using simple mappings, such as a linear transformations~\cite{jha2018elemnet, cao2019convolutional}. Ideal input features are (i) general: they maintain predictive performance across a broad range of materials, (ii) extensible: they can be constructed from readily available data sources or simple calculations, and (iii) interpretable: they can provide insights into underlying physics or correlations which aid in predicting the material property of interest~\cite{kalidindi2020feature}. 

The discovery of features meeting the aforementioned three criteria remains one of the primary challenges of scientific machine learning (SciML) and represents a principal roadblock on the path toward interpretable physics-informed models~\cite{ouyang2018sisso, seko2017representation}. Features for materials informatics models are often engineered to encode one or more material properties, including those derived from chemical composition, topology, electronic behavior, or structural fingerprints~ \cite{mikulskis2019toward}. The most commonly used features are composition-based vectors constructed from properties of the constituent elements of the material~\cite{murdock2020domain, seko2014machine} such as those included in the Magpie feature set~\cite{ward2016general}, which provides descriptive statistics such as minimum, maximum, mean, and range of a set of tabulated element properties for a given chemical composition. Magpie has been used for building successful models across a broad range of materials classes \cite{ling2017high, cao2019convolutional, stanev2018machine} and serves as the foundation for the widely-used Matminer suite of open-source materials informatics tools~\cite{ward2018matminer}. While the standard Magpie library enables simple human-interpretable featurization of chemical composition, its strong reliance on elemental composition generally leads to a lack of robust encoding of underlying interactions beyond those of the material's constituent elements. Featurization based on elemental composition is increasingly being augmented with geometric and topological information such as crystal structure, lattice constants, bond types, and graph-based molecular representations~\cite{emery2017high, tetko2016development, venkatraman2018predicting, seko2017representation, sivaraman2020machine, xie2018crystal},  complex spatial descriptors such as the many-body tensor representation~\cite{huo2017unified},  and string-based molecular encodings such as SMILES~\cite{weininger1988smiles}, BigSMILES~\cite{lin2019bigsmiles}, and SELFIES~\cite{krenn2020self}. Structure-based encodings are widely utilized for featurization in academic studies~\cite{jackson2019recent, kearnes2016molecular, gu2019machine}, but obtaining details about crystallinity or molecular structure for large sets of candidate materials can be infeasible in practice if the structure of materials in the design space is not well known. Moreover, graph-based topological featurization methods are inherently difficult to implement for materials with poorly-defined structures such as glasses and multi-principal element alloys.

Recent research efforts suggest that simple composition-based features may be used to build predictive SciML models without sacrificing human interpretability or relying on prior knowledge about material structure~\cite{seko2017representation, perim2016spectral, cheney2007prediction}. Often this is accomplished through the use of feature engineering, the refinement of model inputs by pruning of existing features or construction of new features from simpler base features, which enables physics-based mapping of model inputs to target properties through the use of symbolic regression~\cite{ouyang2018sisso, kalidindi2020feature}. Similar methods have been demonstrated for building interpretable classification models in health care and criminal justice~\cite{rudin2018optimized, zeng2015interpretable, angelino2017learning}, but quantitative comparisons between the performance of black box models and interpretable models enabled by symbolic regression have not been widely documented for SciML applications~\cite{ouyang2018sisso,xiang2021physics}.

In this study we investigate the use of empirically-derived feature vectors for constructing interpretable predictive models. We screen models by their ability to extrapolate to new clusters in the input design space using leave-one-cluster-out (LOCO)~\cite{meredig2018can} cross-validation (CV), a specific implementation of leave-group-out (LOG)~\cite{lu2019error} and other extrapolation-based CV techniques~\cite{roberts2017cross} (Figure~\ref{fig:workflow}). We compare the extrapolation performance of linear models which rely on a single interpretable feature to that of black box random forest and neural network models which utilize up to $10^{2}-10^{3}$ different input features. As a case study, we examine model extrapolation on a set of 9 open datasets: (i) 3 materials datasets with Magpie featurization, (ii) 3 materials datasets without Magpie featurization, and (iii) 3 non-materials datasets in the physical sciences which are used to test generality of the results. The study highlights important considerations for balancing model performance and interpretability and demonstrates the potential for building interpretable single-feature linear models with extrapolation performance that is comparable to that of black box algorithms in many SciML problems.


\section*{Methods}

\begin{table}
\centering
\caption{Details of test datasets, including the target variable, target description, regression inputs, type of data (experimental or computational), dataset shape before feature engineering (number of samples, number of columns), and dataset source.}

\begin{tabular} {>{\raggedright}p{0.13\linewidth}>{\raggedright}p{0.18\linewidth}>{\raggedright}p{0.25\linewidth}p{0.05\linewidth}p{0.09\linewidth}p{0.12\linewidth}}
\toprule
Target & Description & Inputs & Type & Shape &Source\\
\midrule

melting temp &
melting temperature of $A_{x}B_{y}$ compounds in~K &
Magpie chemistry (e.g. atomic numbers, atomic weights, valences) &
exp. &
(243, 101) &
Ref.~\cite{seko2014machine} \\

 & & & & & \\
 
bulk modulus &
bulk modulus of M$_{2}$AX compounds in~GPa &
Magpie chemistry (e.g. lattice constants, atomic numbers, valences) &
comp. &
(223, 74) &
Matminer~\cite{m2axurl} \\

 & & & & & \\
 
band gap &
electronic band gap of double perovskites in~eV &
Magpie chemistry (e.g. atomic numbers, atomic weights, valences) &
comp. &
(1306, 69) &
Matminer~\cite{doubleperovskitesgapurl} \\

 & & & & & \\
 
heat capacity &
heat capacity of organic molecules at 298.15~K in cal$\cdot$mol$^{-1}\cdot$K$^{-1}$ &
non-Magpie chemistry (e.g. rotational constants, HOMO, LUMO, ZPVE) &
comp. &
(1331, 11) &
DeepChem~\cite{qm9url} \\

 & & & & & \\
 
compressive strength &
compressive strength of concrete formulations in~MPa &
non-Magpie formulations (e.g. water, fly ash, cement content) &
exp. &
(985, 9) &
UCI~\cite{concreteurl} \\

 & & & & & \\
 
formation energy & 
formation energy of transparent conductors in~eV per atom &
non-Magpie chemistry (e.g. lattice vector angles, bandgap, percent aluminum) &
comp. &
(2137, 13) &
Kaggle~\cite{tcourl} \\

 & & & & & \\
 
fish weight &
weight of fish in g &
physical measurements (e.g. length, height, width) &
exp. &
(158, 6) &
Kaggle~\cite{fishurl} \\

 & & & & & \\
 
airfoil sound &
intensity of sound in decibels created by airfoil flight &
 physical measurements (e.g. airfoil angle, velocity, chord length) &
 exp. &
 (1497, 5) &
 UCI~\cite{airfoilurl} \\
 
 & & & & & \\
 
abalone rings &
number of rings (corresponding to age) in an abalone shell &
physical measurements (e.g. length, diameter, weight) &
exp. &
(4175, 8) &
UCI~\cite{abaloneurl} \\

\bottomrule
\end{tabular}
\label{table:datasetstable}
\end{table}

\begin{figure}
\centering
\includegraphics[width=\linewidth]{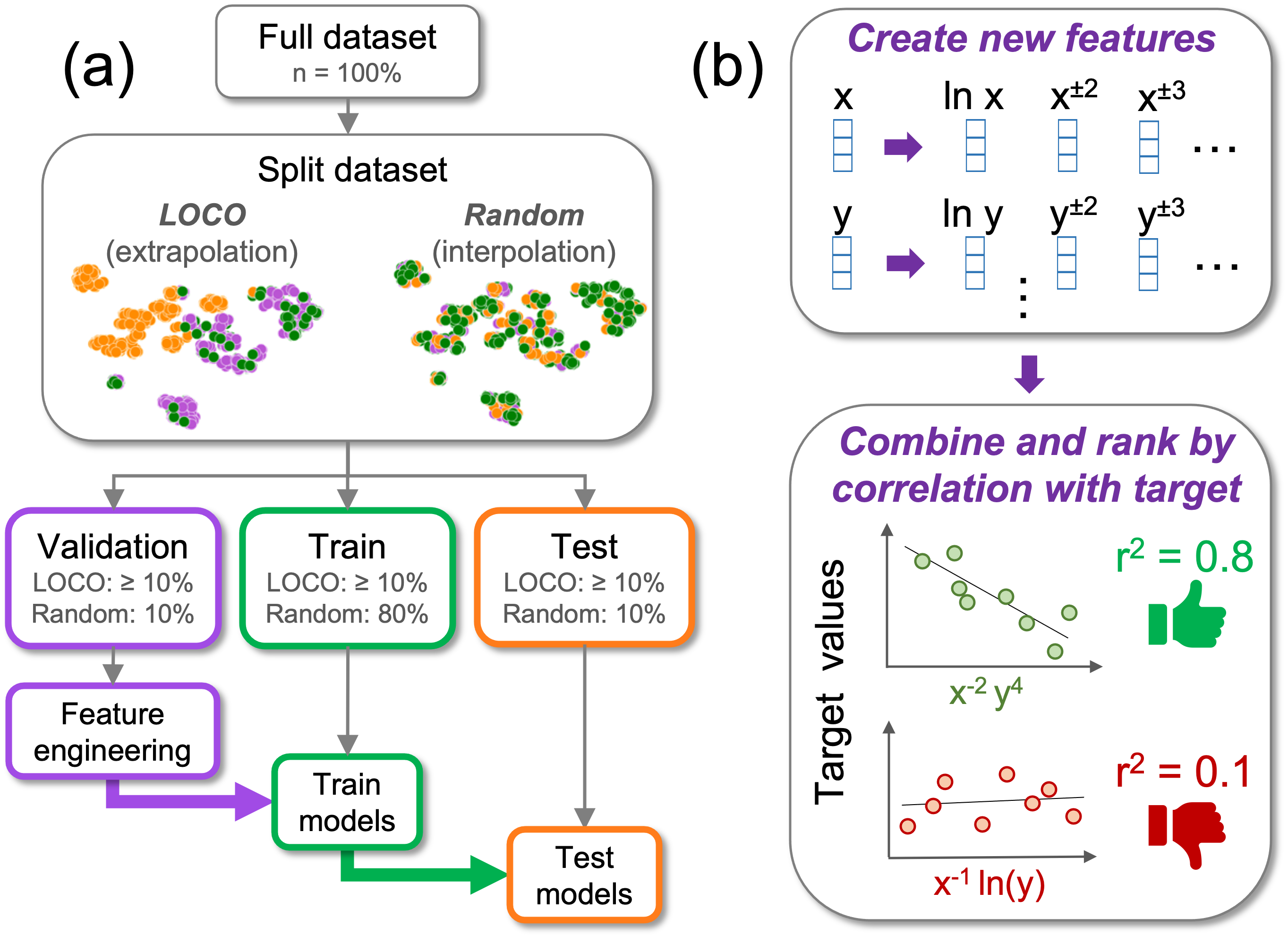}
\caption{(a) Schematic of experimental workflow in which each test dataset was split using LOCO CV for testing extrapolation and random CV for testing interpolation. The validation set (purple) was used for engineering of new features, the training set (green) was used for model fitting, and the test set (orange) was used for evaluating model performance. For LOCO CV, each of the train, test, and validation sets contained at least 10\% of the dataset. For random CV, 80\% of the dataset was using for training, 10\% for validation (feature engineering), and 10\% for testing model performance. (b) Schematic of feature engineering process in which additional features were created from existing features (top panel), combined multiplicatively with other features, and sorted by their Pearson $r^{2}$ correlation to the target values in the validation set (bottom panel). }
\label{fig:workflow}
\end{figure}

Benchmark datasets were selected to span a broad range of materials informatics problems, using both Magpie- and non-Magpie-featurized materials, as well as non-materials scientific datasets to test generality of the results beyond materials. One variable in each dataset was selected as the regression target to simulate a real-world scientific modeling problem. Target variables were selected to sample a broad range of material properties, including electronic, mechanical, and thermodynamic properties, and to ensure diversity in target value distributions across each dataset. 

Prior to partitioning into cross-validation (CV) sets, datasets were cleaned by removal of rows which contained at least one value which was greater than 3 interdecile ranges away from the median value of the column, removal of duplicate rows and columns, removal of non-numeric columns,  removal of rows containing null values, and removal of columns with less than 5 unique values. The 9 cleaned benchmark datasets, along with a description of each target variable, dataset size, and data source, are shown in Table~\ref{table:datasetstable}. Histograms of each target value are shown in Figure~S1 where each panel corresponds to one test dataset and is labeled by the name of the target variable in the dataset. Three materials datasets (colored green in Figure~S1) were featurized based on chemical formula using the \emph{element\_property}, \emph{BandCenter}, and \emph{AtomicOrbitals} featurization methods in Matminer. Features generated from Matminer which contained non-numeric values or constant values across all rows were removed, resulting in slightly different sets of Matminer features in each dataset.

To prevent data leakage during feature engineering and model testing, each dataset was partitioned into 3 splits: train (for training predictive models), test (for evaluating model performance), and validation (for assessing the quality of engineered features), as shown in Figure~\ref{fig:workflow}(a). For each dataset, data partitioning was repeated 10 times for each of 2 CV strategies: leave-one-cluster-out (LOCO CV), to simulate model extrapolation (Figure~S2), and random CV, to simulate model interpolation (Figure~S3), resulting in 20 total unique train-test-validation splits per dataset. For LOCO CV, splits were generated following the method of Meredig et al. ~\cite{meredig2018can}. Standard normalization of input features was performed, the data was shuffled, k-means clustering with random initialization was performed where the number of clusters was selected randomly between 3 (the minimum number of clusters required for train, test, and validation sets) and 10 (the largest traditional choice for clusters in cross-validation), and each cluster was randomly assigned to a train, test, or validation set such that no set contained less than 10\% of the entire dataset. For random CV, the dataset was divided randomly so that 80\% was used for training, 10\% was used for testing, and 10\% was used for validation. The uniqueness of each CV split was verified to ensure that no two splits contained identical partitioning of samples.

After datasets were partitioned into train, test, and validation sets, feature engineering was performed with the goal of creating interpretable input features built from the set of existing base features, as described in Figure~\ref{fig:workflow}(b). To enable a variety of interpretable functional forms for engineered features, base input features were raised to varying powers: from feature $x$, the features $x^{-1}$, $x^{\pm 1/2}$, $x^{\pm 1/3}$, $x^{\pm 1/4}$, $x^{\pm 2}$, $x^{\pm 3}$, $x^{\pm 4}$, and $\ln x$ were generated when possible (for example, $x^{-1}$ was not generated for feature $x$ when feature $x$ contained values equal to 0). Next, between 2 and 5 of the newly-constructed features were selected at random. The selected features were multiplied together to generate a new feature. This process was repeated $5 \times 10^{5}$ times at each CV split for each dataset, and each resulting feature was ranked by its Pearson $r^{2}$ correlation to the target values in the validation set. The engineered feature which exhibited the highest $r^{2}$ correlation to the target values of the validation set at a given CV split was selected as the best engineered feature (BE) for that CV split. While recent efforts have demonstrated more advanced feature engineering methods for materials informatics~\cite{ouyang2018sisso, xiang2021physics}, the random search approach described here was used to demonstrate how feature engineering can be performed using a simple procedure without the need for dedicated software packages or computationally-expensive genetic algorithms. No constraints were applied to enforce dimensionality of the resultant features.

After the BE feature was selected for each CV split, it was used for fitting single-feature linear regressions which were compared to black box models in both interpolative and extrapolative regimes. At each CV split, 3 regression models were compared: a random forest, a neural network with stochastic gradient-descent optimizer, and a linear regression, all implemented using default Scikit-learn methods (\emph{RandomForestRegressor}, \emph{MLPRegressor}, and \emph{LinearRegression}, respectively) with default hyperparameters (100 estimators and no maximum tree depth for the random forest, and 1 hidden layer with 100 neurons, ReLU activation, Adam solver, and 200 maximum iterations for the neural network)~\cite{pedregosa2011scikit}.  Default hyperparameters were used to investigate the performance of models with standard architectures for both interpolation and extrapolation tasks, as optimization of hyperparameters often yields improvements in model interpolation at the potential expense of performance in extrapolative settings,  which may lead researchers to overestimate model utility for real-world problems~\cite{meredig2018can}.

Before fitting the models, each dataset was scaled using the Scikit-learn \emph{RobustScaler} method. Model performance was quantified using non-dimensional model error (NDME)~\cite {liu2021dielectric}, calculated as the ratio between root-mean-square error (RMSE) between predicted values and ground truth values in the test set, and the standard deviation in the ground-truth values of the test set. NDME was used as a performance metric to ensure that model performance could be compared across datasets and target properties with varying units and magnitudes, where NDME of 0 corresponds to a perfect model, and NDME of 1 corresponds to a model with average prediction error which is equal to standard deviation in the ground truth values. Models were evaluated on their ability to predict target values in the test set of each CV split for 3 different featurization strategies: (1) best engineered (BE), where the single best engineered feature at the given CV split was used as the only model input, (2) BE + original, where all the original dataset input columns, in addition to the single best engineered feature at the given CV split, were used as input features, and (3) original, where the original dataset input columns were used as input features.  Covariance confidence intervals of model NDMEs were calculated by excluding extreme model outliers with NDME~>~5.

\begin{figure}
\centering
\includegraphics[width=\linewidth]{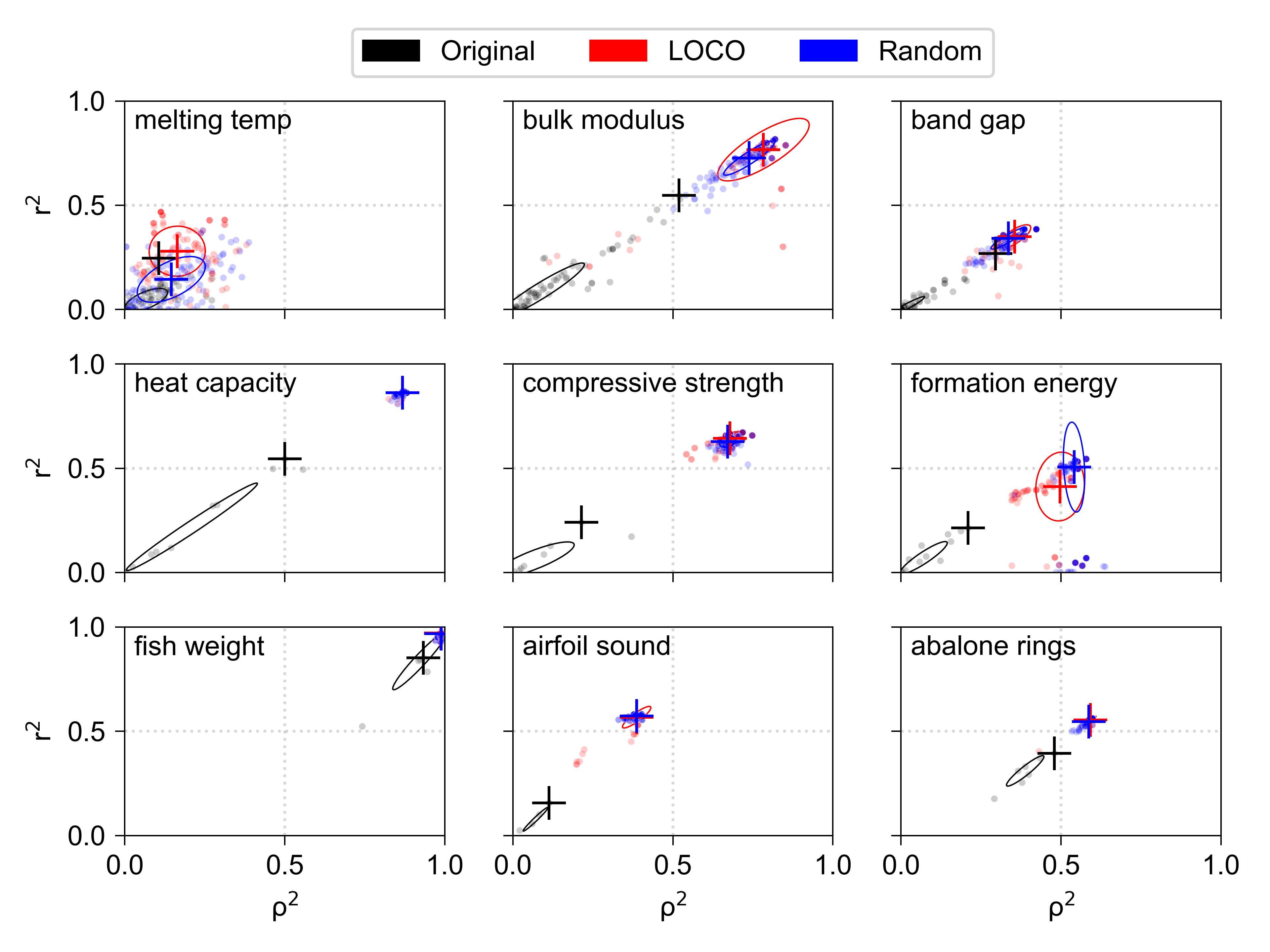}
\caption{Pearson $r^{2}$ (vertical axes) and Spearman $\rho^{2}$ (horizontal axes) correlations between input features and all target values in each dataset (only the validation set was used for feature engineering). Each original input feature for a given dataset is shown as a single black point, while the top 100 engineered features discovered using LOCO CV random CV are shown by red and blue points respectively. Large cross symbols denote the median location of discovered features in the LOCO and random groups (red and blue), and the location of the best feature (feature with the highest $r^{2}$ value) in the original input feature set (black).  Confidence ellipses (1 standard deviation from the median) are shown for Original, LOCO, and Random feature sets.  Feature engineering was performed with the goal of discovering novel features with high $r^{2}$ correlation to the target variable in each dataset. }
\label{fig:correlationscores}
\end{figure}

\section*{Results and discussion}

The primary challenge of creating interpretable physics-informed models lies in the discovery of features which provide a linear mapping between input values and target values. We performed feature engineering with the goal of finding new features which were linearly correlated with the target variable of each test dataset (Figure~S4). Figure~\ref{fig:correlationscores} summarizes the input features associated with each test dataset before and after the feature engineering process. Each panel corresponds to one dataset and is labeled by the target variable name. Each scatter point corresponds to a single input feature, where the original features of each dataset are shown in black, the top 100 features engineered using LOCO CV (top 10 features for each of 10 CV splits) are shown in red, and the top 100 features engineered using Random CV are shown in blue. Each feature is plotted by its Pearson $r^{2}$ correlation with the target variable on the vertical axis, for quantifying linearity with the target, and its and its Spearman $\rho^{2}$ correlation on the horizontal axis, for quantifying monotonicity with the target. Large cross symbols show the median location of engineered points in the LOCO and random groups, and the single feature with the highest $r^{2}$ value in the original feature set. While top features were selected during the feature engineering process based on their $r^{2}$ correlation to the validation set, the correlations shown in Figure~\ref{fig:correlationscores} represent correlations between input features and the target value of the entire dataset. Points which lie near the origin (0, 0) represent features which are generally non-informative for linear models, as they lack linearity and monotonicity with the target variable. Features near the coordinate (0, 1) exhibit high linearity but poor monotonicity, which suggests that high degeneracy in target values may be predicted for a single value of the input feature. Features near coordinate (1, 0) exhibit high monotonicity but poor linearity, which makes them informative inputs for nonlinear regression models. Features near coordinates $(1,1)$ exhibit both high linearity and monotonicity with the target values, which makes them suitable for use in simple linear regression models. 
 
In every dataset, the feature engineering process uncovered new features with $r^{2}$ correlation to the target variable which was higher than that of any of the original features. In 4 of 9 datasets (\emph {heat capacity} \emph{compressive strength}, \emph{formation energy}, \emph{ airfoil sound}), feature engineering resulted in discovery of new features which achieved significant (>100\%) increases in $r^{2}$ correlation to the target, which suggests that simple algebraic manipulation and combination of existing features may be used for creating highly informative input features in many SciML problems. A noteworthy result is that higher correlations between original features and the target variable do not necessarily enable engineering of new features with proportionally higher correlations. For example, the best original input feature for \emph{abalone rings} exhibits $r^{2}$ correlation of ${\sim}0.45$, while that of \emph{compressive strength} exhibits $r^{2}$ of ${\sim}$0.2. However, feature engineering for \emph{compressive strength} results in new features with median $r^{2}$ of ${\sim}0.7$, whereas $r^{2}$ for engineered \emph{abalone rings} features reaches a maximum value at ${\sim}0.6$. The results demonstrate that the return on investment in feature engineering is highly dependent on characteristics of the specific dataset and may be difficult to predict from correlations between the the original input features and the target variable alone.

\begin{table}
\caption{Sample of the best engineered features for each dataset. Each row contains the target variable name, the functional form of the best engineered feature, the Pearson $r^{2}$ correlation between the best engineered feature (BE~max~$r^{2}$) and target values in the full dataset, and the highest $r^{2}$ correlation between any of the original input features and target values in the full dataset before feature engineering (Orig.~max~$r^{2}$).}
{\tabulinesep=1.2mm
\begin{tabu}{p{0.12\linewidth}p{0.54\linewidth}p{0.1\linewidth}p{0.12\linewidth}}
\toprule
Target &
\centering Engineered feature &
BE~max~$r^{2}$ &
\centering Orig. ~max~$r^{2}$\\
\midrule

melting temp &
\centering $ \displaystyle  \frac{\ln (\text{modeColumn}) \cdot \text{rangeMeltingT}^{1/2}} {\text{modeNValence}^{2} \cdot \text{modeSpaceGroupNum}^{1/4} }  $ &
0.47 &
0.25 \\

bulk modulus &
\centering $ \displaystyle  ( \text{minMendeleevNumber} \cdot \text{modeMeltingT} )^{1/2} \cdot \frac { \text{d\_mx}^{2} \cdot \ln(\text{maxCovalentRadius})} { \text{c}^{3} } $ &
0.82&
0.55 \\

band gap &
\centering $ \displaystyle    \text{rangeNUnfilled} \cdot \text{meanNsValence}^{4} \cdot \left( \frac { \text{meanNpValence} } { \text{meanElectronegativity} \cdot \text{meanColumn} } \right)^{3} $ &
0.39 &
0.27 \\

heat capacity &
\centering $ \displaystyle  \frac{\text{zpve}^{1/2} \cdot \text{r2}^{3/4}} {\text{A}^{1/3}}  $ &
0.87 &
0.55 \\

compressive strength &
\centering $ \displaystyle  \frac{\text{cement}^{1/2} \cdot \text{day}^{1/4}} {\text{fineAggregate}^{1/2} \cdot \text{water}^{3/2}} $ &
0.67 &
0.24 \\

formation energy &
\centering $ \displaystyle \text{percent\_atom\_al}^{1/2} \cdot \text{percent\_atom\_in}^{1/4} \cdot \left( \frac{ \text{lattice\_vector\_3\_ang} } { \text{bandgap\_energy\_ev} \cdot \text{lattice\_angle\_beta} } \right)^{1/3} $ &
0.55 &
0.21 \\

fish weight &
\centering $ \displaystyle \ln(\text{width}) \cdot \text{length2}^{2} \cdot \left( \frac{\text{height}} {\text{length3}} \right)^{1/2} $ &
0.98 &
0.85 \\

airfoil sound &
\centering $\displaystyle  (\text{chordlength\_m} \cdot \text{displacentthickness\_m} )^{1/2}  \cdot  \text{frequency\_hz}^{2/3}$ &
0.58 &
0.16 \\

abalone rings &
\centering $ \displaystyle \frac{\text{whole\_weight} \cdot \text{shell\_weight}^{1/3}}{\text{shucked\_weight}} $ &
0.56 &
0.39 \\

\bottomrule

 \end{tabu}}
\label{table:featuretable}
\end{table}

\begin{figure}
\centering
\includegraphics[width=\linewidth]{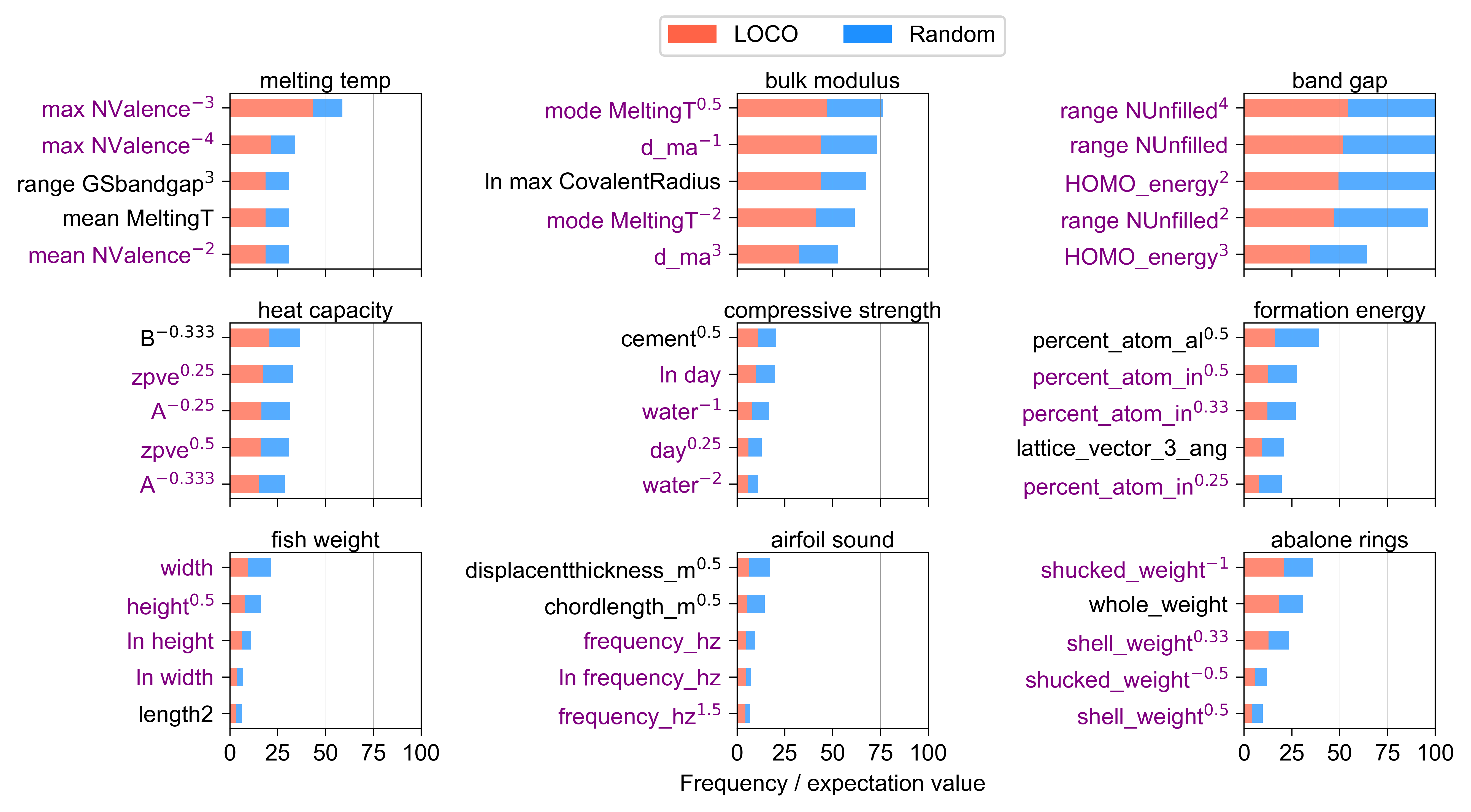}
\caption{Prominence of variables in top-performing engineered features for a given dataset. Each panel corresponds to a single dataset and shows the frequency of CV splits in which each variable appeared in the top 10 engineered features for that dataset, divided by the expectation value of the frequency if engineered features had been selected at random.  Purple variable labels correspond to variables which occur more than once in the top 5 variables.}
\label{fig:featureprominence}
\end{figure}

A sample of the best engineered feature discovered per dataset (that which exhibited the highest $r^{2}$ correlation to the full dataset, after being evaluated only on its correlation with the validation set during feature engineering) is shown in Table~\ref{table:featuretable}. The variable names present in each feature represent the corresponding column names in the original datasets (including the feature names created from Magpie featurization in the first 3 test datasets). Variable names follow the standard Matminer naming convention: minimum (min), maximum (max), range, mean, mode, and average deviation (avgdev). The Pearson $r^{2}$ correlation between the engineered feature and the target variable is shown in the BE $r^{2}$ column, while the highest $r^{2}$ correlation between the target variable and any original input features is shown in the \emph{Orig.} $r^{2}$ column. Both $r^{2}$ values were calculated using the full dataset, even though features were engineered based on their correlation to the validation set only.

The human-readability of each engineered feature enables rapid identification of the input variables which are predictive of the given target variable, in addition to whether input variables are directly or inversely proportional to the target variable, and whether they exhibit higher correlation to the target variable when raised to a specific power or after taking the natural logarithm. For example, \emph{heat capacity} is highly correlated ($r^{2} = 0.87$) with the square root of zero-point vibrational energy (zpve) divided by the cube root of $A$, a rotational constant. This is consistent with general thermodynamic considerations, as vibrational and rotational degrees of freedom give rise to heat capacity~\cite{sebbar2002structures, grev1991concerning}, and heat capacity is proportional to a factor with dependence on zpve~\cite{gomaa2019thermal}. Fish weight is strongly predicted ($r^{2} = 0.98$) by the square of the \emph{length2} property multiplied by factors related to width and the ratio of height to \emph{length3},  which differs from the basic intuition that weight scales linearly with volume,  which in a first-order approximation is the simple product of length, width, and height. The feature for predicting the \emph{compressive strength} of concrete includes roughly half of the variables available in the original dataset, which suggests that several ingredients in the concrete formulation (water, cement, aggregates) are significantly more important for predicting compressive strength than others (\emph{fly ash}, \emph{furnace slag}, and \emph{superplasticizers}) which are not present in the engineered feature. These insights are not easily obtained through the use of black box models, which encode the relationships between different input variables using large arrays of coefficients that are generally not human-readable due to their high dimensionality.

The results in Table~\ref{table:featuretable} demonstrate how model complexity may be offloaded from the internal model structure to the input features. In black box models, the relationships between inputs are generally represented by complex data structures (tree ensembles in random forests, hidden layer node interconnections in neural networks), which are difficult to interpret by human domain experts.  Using symbolic regression for feature engineering, these relationships can be represented in human-readable form\cite{angelino2017learning}.  The human-readable features exhibit higher correlations with the target variable than any of the original features in the dataset,  which unlocks the potential for using simple linear regression models instead of black box algorithms to achieve the same predictive power. Transfer of the model complexity to the input features significantly enhances model portability, as the resulting linear models can be fully represented by just 2 coefficients: slope and intercept, which makes them easy to deploy in a broad range of settings without the need for specialized machine learning experts.

Figure~\ref{fig:featureprominence} shows the 5 variable functional forms which were most frequently included in the top 10 engineered features per CV split of a given dataset. The horizontal axis is scaled by the expectation value of the frequency of each variable in the top 10 engineered features if it had been selected randomly. The functional form of each variable, i.e. whether they were raised to powers or a natural logarithm was taken, is shown in each variable label.  On average,  ${\sim}$3.7 of the top 5 variables appeared in the top 5 variables more than once in different functional forms (shown by an average of 3.7 purple variable labels per panel), which suggests that a small number of informative features were used frequently during the feature engineering process due to their high correlation to the target value compared to other variables in the dataset. The results enable key diagnostics about which input variables, and which functional forms of those input variables, contributed to informative inputs for feature engineering. Unlike standard feature importance rankings which are commonly reported for ensemble regression methods~\cite{pedregosa2011scikit} or make use of game-theoretic formulations such as Shapley~\cite{kumar2020problems}, the methodology demonstrated here enables ranking of the important input features while additionally highlighting the corresponding functional forms (i.e.  power, logarithm) which yield highest correlations to target values, and highlights the resiliency of certain input variables to remain informative over different regions (CV splits) of the dataset.

It is worth noting that variables in the first 3 datasets in Figure~\ref{fig:featureprominence} generally appear in higher frequency than those in subsequent datasets.  The large number of Magpie columns in the first 3 datasets (see dataset size in Table~\ref{table:datasetstable}) lowers the expectation value that a single variable will appear in high-performing engineered features across multiple CV splits,  which increases the value of Frequency~/~expectation plotted on the horizontal axis.  The occurrence of some properties in multiple functional forms,  particularly \emph{NValence} in the \emph{melting temp} dataset and \emph{range NUnfilled} in the \emph{band gap} dataset,  attests to the powerful utility of those properties for predicting the given target variable across a broad range of extrapolative CV splits.

\begin{figure}
\centering
\includegraphics[width=\linewidth]{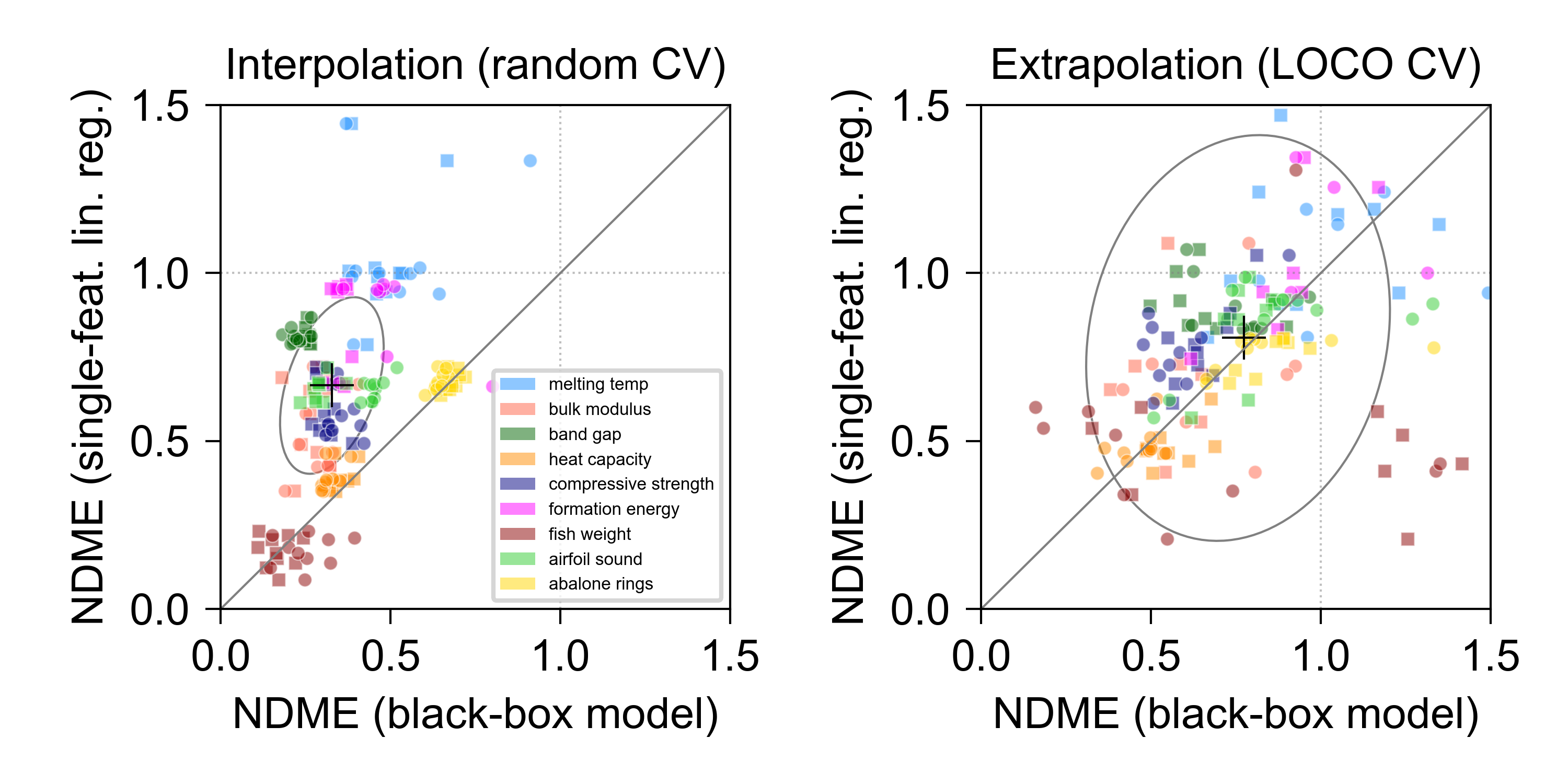}
\caption{Model performance during interpolation (random CV, left panel) and extrapolation (LOCO CV, right panel). NDME of single-feature linear models (a linear regression fitted using the best engineered feature discovered at the given CV split as its only input feature) is plotted on the vertical axis against NDME of black-box models using all original input features as inputs on the horizontal axis. Each point corresponds to the NDME at one CV split and is colored by dataset. Square points correspond to RF models and circular points correspond to NN models. The large cross symbol denotes the median location of all points. The covariance confidence ellipse denotes the location 1 standard deviation away from the median value. The solid line at $y = x$ represents the point at which a single-feature linear model and a black box model exhibit equal performance at a given CV split and given dataset. Dotted lines show NDME~=~1, the point at which model prediction error is equal to the standard deviation in ground truth values.}
\label{fig:extrapolationandinterpolation}
\end{figure}

To investigate the trade-off between model performance and interpretability, we compared the prediction error of single-feature linear models to that of black box models in both interpolation and extrapolation problems. The BE feature at each CV split was used as the input for a single-feature linear regression. The prediction accuracy of the linear regression was compared to that of two black box algorithms: one random forest (RF) and one feed-forward back-propagating neural network (NN). Both black box algorithms were fitted using all original input features as inputs. Results from the model comparison are shown in Figure~\ref{fig:extrapolationandinterpolation}. 

For interpolation problems (random CV, left panel), the median NDME across all CV splits was ${\sim}$0.67 for single feature linear models and ${\sim}$0.33 for black box models, and black box models outperformed linear regressions in 87\% of CV splits. This is consistent with the understanding that complex algorithms can effectively capture subtleties of the response surface in the immediate neighborhood of training data with arbitrary precision~\cite{agrawal2019deep}. In three datasets (\emph{fish weight}, \emph{heat capacity}, and \emph{abalone rings}), the linear models performed comparably to the black box models (median NDME for linear models was similar to that of black box models). In the \emph{band gap},  \emph{formation energy}, and \emph{melting temp} datasets, linear models exhibited NDME which was roughly twice as high as that of corresponding black box models. These datasets exhibited some of the lowest BE $r^{2}$ values of any datasets in Table~\ref{table:featuretable}, which supports the intuition that higher $r^{2}$ correlations between engineered features and target variables enable the construction of higher-performing linear models. 

In extrapolation problems (LOCO CV, right panel), the performance difference between single-feature linear regressions and black box models was much less pronounced. The median NDME for linear models, ${\sim}$0.81, was an average of ${\sim}$5\% higher than that of the NDME for black box models (${\sim}$0.77). Black box models outperformed linear regressions in 60\% of CV splits, with roughly ${\sim}$40\% of the area of the covariance confidence ellipse, shown by the gray curve, lying inside the region below the $y=x$ line, which suggests that the performance difference between linear models and black box models is highly dependent on the particular dataset and extrapolation region (CV split), and difficult to predict and generalize across different datasets. For extrapolation in the \emph{fish weight} and \emph{heat capacity} datasets, single-feature linear regressions almost always outperformed black box models. The engineered features discovered for those datasets exhibited very high median correlations to the target values ($r^{2} > 0.85$, Figure~\ref{fig:correlationscores}), which indicates that the $r^{2}$ correlation between input features and target variables may serve as a rough guide for predicting whether single feature linear models will outperform black box models in extrapolation problems. 

We also examined the effect of including BE features as inputs to the black box models. For each dataset, the performance of linear regression models was compared to that of RF and NN models using 3 different featurization strategies: (1)~original: the model was trained using only the original input features in the dataset, including those calculated using Magpie, (2)~best engineered~(BE): the model was trained using the single best engineered input feature for the given CV split as ranked by its $r^{2}$ correlation to the validation set, and (3)~BE~+~original: the model was trained using all original input features in addition to the single best engineered feature. Model performance was quantified by the median NDME across all train-test splits for a given dataset and extrapolation or interpolation task (Figure~S5).

In all cases, the best performing model configuration for a given dataset exhibited lower NDME when performing interpolation than when performing extrapolation, which is consistent with the common understanding that extrapolation outside of the range of training data is more difficult than interpolation of data which was used during the training process~\cite{meredig2018can}. Model extrapolation performance either improved or stayed unchanged upon addition of the single best engineered feature in the majority of all test cases, which indicates that the feature engineering process for addition of input variables generally provides performance benefits without risk of decreasing the prediction accuracy. This result suggests that when time and computational resources allow it, the feature engineering process should almost always be used prior to model training. In 4 of 9 test datasets (\emph{bulk modulus}, \emph{heat capacity}, \emph{fish weight}, and \emph{abalone rings}), single-feature linear models achieved extrapolation performance which was comparable (within error bars, which represent the standard deviation in NDME values cross the CV splits) to that of the best-performing black box models. 

The results are noteworthy because they demonstrate that the generally-assumed trade-off between model interpretability and performance is often overstated, especially in extrapolation contexts. Linear regressions which utilize a single input feature in human-readable functional form can often extrapolate just as well as black box models which utilize $10^{2}-10^{3}$ inputs. The linear models exhibit superior portability, decreased computational overhead, and improved ease of use by non-experts over black box models, which highlights their potential for deployment in edge computing applications and other resource-constrained environments without the need for expert operators. 

\section*{Conclusions}

Selection of predictive models for SciML problems is often driven by the assumption that an implicit trade-off exists between model performance and complexity. We investigated this trade-off by engineering human-readable input features for a set of 9 open datasets using a simple random search algorithm, and then compared the performance of single-feature linear regressions to that of black box random forest and neural network machine learning models in interpolative and extrapolative settings across different SciML tasks.  For interpolation tasks, the linear models exhibited average prediction error roughly twice as high as that of the black box models, and outperformed black box models in just 13\% of cross-validation splits. However, when extrapolating to new regions of the dataset which were not present in the training set, linear models exhibited average prediction error ${\sim}$5\% higher than that of black box models, and outperformed black box models in ${\sim}$40\% of cross-validation splits, while enabling faster model training times and improved model portability. The use of human-readable input features provided insight into the functional forms of underlying variables which were most predictive of a given set of target values. The results suggest that extrapolation in many SciML problems may be performed using simple linear models which maintain human interpretability without sacrificing performance. Increased use of simple interpretable models may help unlock the full potential of machine learning for light-weight edge computing, rapid model portability, and generation of data-driven insights which augment existing human domain knowledge.

\section*{Data availability}
Data and code related to this paper are available at https://github.com/CitrineInformatics-ERD-public/linear-vs-blackbox.

\section*{Acknowledgement}
This research was developed with funding from the Defense Advanced Research Projects Agency (DARPA) and material is
based upon work supported by the United States Air Force under Contract No. FA8650-20-C-7002. Any opinions, findings and
conclusions or recommendations expressed in this material are those of the author(s) and do not necessarily reflect the views of
the United States Air Force.

\bibliographystyle{unsrt}
\bibliography{references}

\includepdf[pages=-]{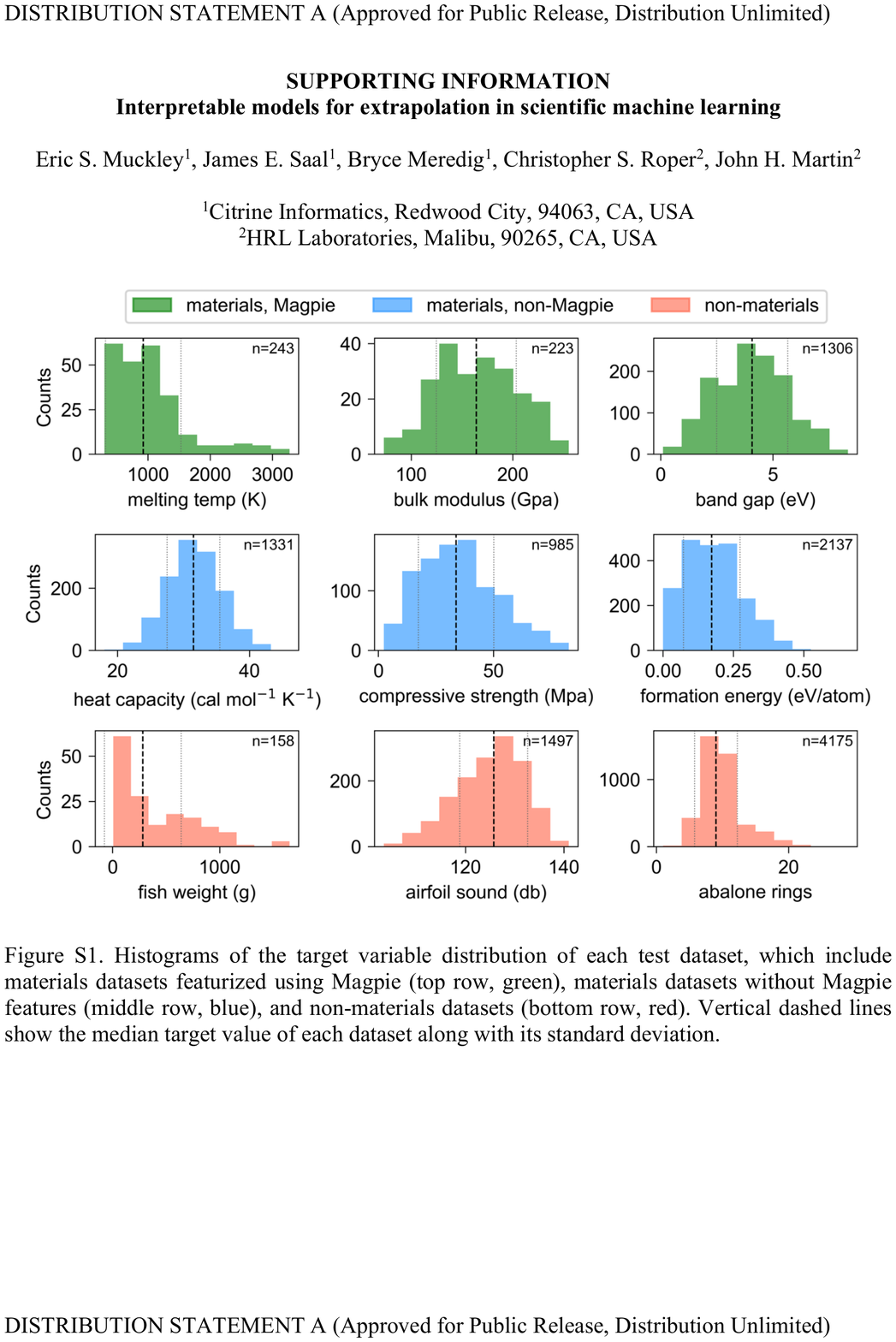}

\end{document}